\documentclass{vgtc}                          

\graphicspath{{figures/}{pictures/}{images/}{./}} 
\RequirePackage[svgnames]{xcolor}

\usepackage{subcaption}

\usepackage{times}                     

\usepackage{tabu}                      
\usepackage{booktabs}                  
\usepackage{lipsum}                    
\usepackage{mwe}                       

\usepackage{mathptmx}                  

\usepackage{amsmath}
\usepackage{amssymb}
\usepackage{tikz}

\newcommand{\norm}[1]{\left\| #1 \right\|}

\onlineid{1273}

\vgtccategory{Research}

\vgtcinsertpkg

\title{Multi-Focus Probes for Context-Preserving Network Exploration and Interaction in Immersive Analytics}

\author{Eric Zimmermann\thanks{e-mail: e.zimmermann@uni-rostock.de}\\ %
        \parbox{1.4in}{\scriptsize \centering University of Rostock, \\ Germany} %
\and Stefan Bruckner\thanks{e-mail: stefan.bruckner@uni-rostock.de}\\ %
        \parbox{1.4in}{\scriptsize \centering University of Rostock, \\ Germany}}

\teaser{
  \centering
  \includegraphics[width=\linewidth]{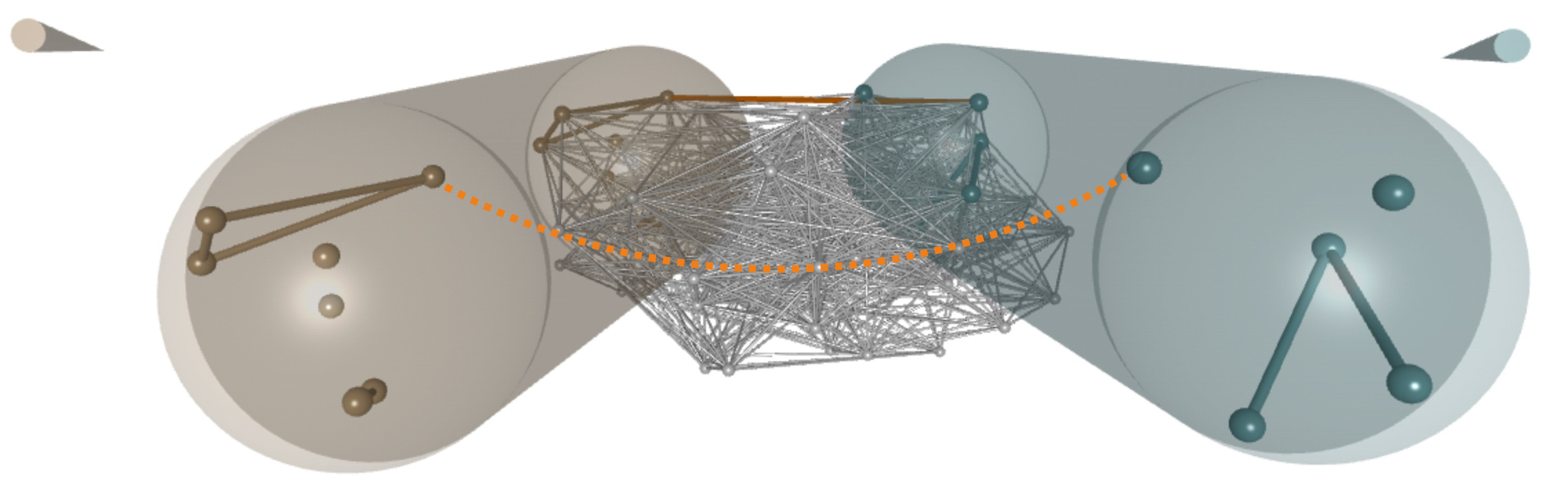}
  \caption{Multiple focuses set as colored 2-spheres upon a graph with their respective colored contents (induced subgraphs) presented in front of the user for inspection, editing, and interaction. The recently added link (highlighted in orange in the back) was created as a local operation joining the two respective nodes in the subgraphs. This link is not represented in either of the subgraphs but instead indicated by the dotted orange line for visual reference. The subgraphs are connected with the probes via tunnels. Additionally, colored cones (on the very left and right) point toward the probes located inside the graph.}
  \label{fig:teaser}
}

\abstract{
Immersive visualization of network data enables users to physically navigate and interact with complex structures, but managing transitions between detailed local (egocentric) views and global (exocentric) overviews remains a major challenge. We present a multi-focus probe technique for immersive environments that allows users to instantiate multiple egocentric subgraph views while maintaining persistent links to the global network context. Each probe acts as a portable local focus, enabling fine-grained inspection and editing of distant or occluded regions. Visual and haptic guidance mechanisms ensure context preservation during multi-scale interaction. We demonstrate and discuss the usability of our technique for the editing of network data.
} 


\begin{document}


%
%
%
%
%
\firstsection{Introduction}
\maketitle

Analyzing large networks in immersive virtual environments offers an engaging 3D experience and potential insight gains, but it also introduces significant usability challenges~\cite{McCrae:2010}. Prior studies have shown that while users can better understand certain graph structures in VR, they often struggle with basic tasks like selecting distant nodes or navigating complex structures without losing their sense of orientation and context~\cite{Drogemuller-2020-EVR,Sorger-2021-ENE}. A recent survey of immersive network analysis identified maintaining spatial orientation in large graphs, managing cognitive load when handling multiple regions of interest, and balancing detailed views with a holistic network understanding as key challenges in the field~\cite{Joos-2025-VNA}. Traditional 2D graph interfaces address the focus+context problem through techniques like overview+detail or fisheye views, but translating these to an embodied 3D setting is non-trivial. For instance, providing an overview of the entire network in VR (such as a World-in-Miniature model~\cite{Pivovar:2022}) can help preserve context, yet switching attention between the overview and the full-scale graph may itself impose cognitive load. Designing interaction techniques that let users examine and edit specific parts of a large graph in situ while keeping their place and perspective in the overall structure remains an open research problem.

In this paper, we propose Multi-Focus Probes, an immersive focus+context technique for network data that allows users reach distant and/or occluded nodes in a virtual graph by placing interactive probes on them. Each probe captures the subgraph within its radius and instantly presents that content in front of the user as a movable, scaled focus view. This approach is inspired by the concept of an endoscope in medicine, enabling inspection and manipulation of hard-to-reach areas through a minimal opening. By bringing chosen sub-networks within arm's reach, the user can perform local edits (such as adding or removing links, or adjusting node positions) directly on the focus views and propagate those changes back to the full graph. The probes serve not only as selection lenses but also as anchors for navigation and even global graph deformation: users can ``teleport'' their viewpoint to a probe's location or apply spatial distortions to the graph via the probe while preserving the overall context. To maintain the spatial relationship between focus and context despite these manipulations, we integrate visual and haptic guidance cues (e.g., connecting lines or tunnels and directional indicators) that always point the user to each probe’s original location in the global graph.

\section{Related Work}

\textbf{Selection and Navigation in VR:} Selecting and manipulating objects in VR has been widely studied, as it forms the basis of more complex analysis tasks~\cite{Argelaguet:2013, Bergstroem:2021}. Common selection techniques like ray-casting involve pointing a controller or hand toward a target~\cite{Pfeuffer:2017}, which works well at moderate distances but can become imprecise for very far or completely occluded targets~\cite{Nukarinen-2018-ERC}. Variations such as the extended arm metaphor allow users to reach beyond their physical arm length into a scene~\cite{Poupyrev:1996, Dai:2025}, overcoming distance limits at the cost of precision~\cite{Dai:2025}. For occluded targets, specialized techniques have been proposed: for example, the Magic Ball approach encloses hidden objects in a transparent sphere to reveal them as a 3D mini-map for selection~\cite{Yu:2020}. Navigation can be considered as an option to support the selection task in form of decreasing the distance between the user and the object \cite{Dai:2025} with different approaches like Point-Tugging or Arm-Cycling~\cite{Coomer:2018} or teleportation~\cite{Bozgeyikli:2016}. Our probe-based selection builds on these ideas by letting users select far-away nodes via ray-casting without directly selecting targets but enclosing them in a radial region to overcome the imprecise control found for ray-casting or extendable arm approaches. Further, we incorporate focus-related navigation which allows to travel directly to the locations of the probes and graph manipulations.

\textbf{Focus+Context and Orientation Aids}: Maintaining an overview while examining details is a classic challenge in visualization. In 2D, solutions include structure-aware fisheye views for large graphs~\cite{Wang-2019-SFV} and interactive lenses for multivariate graph exploration~\cite{Tominski:2024}. In immersive environments, this problem is exacerbated by the lack of fixed external frames of reference. Researchers have explored using the lens metaphor to offer alternative representations even in combination of two lenses~\cite{Kluge:2020} or in terms of direct spatial manipulation of 3D data analysis dealing with surfaces or streamlines~\cite{Mota:2018}. Alternatively, space-folding techniques with local focus views were used \cite{Elmqvist-2008-Melange, Reiske-2023-MFQVR}, for instance to allow the interaction with genome-scale data. These methods keep the overall context-relation, while zooming into details could also be supported by an overview in VR using techniques like World-in-Miniature (WIM)~\cite{Pivovar:2022}. A property of all these approaches is that they allow the user to seamlessly change states between focus and context building a bridge in terms of spatial directness in visualization tasks~\cite{Bruckner:2019}. Consequently, these focuses can be utilized as interaction proxies or portals for exploration \cite{Satriadi-2020-MAM} and remote manipulation~\cite{Stoev:2002, Han:2022, Dai:2025}. In addition, aids can assist the user in keeping orientation in a 3D scene populated by visualizations of data~\cite{McCrae:2010}. In virtual environments it might be more appropriate to consider attention guidance with least possible distraction to outweigh for instance degradation of immersion and presence~\cite{Grogorick:2020} or offer guidance in interactive exhibition scenarios~\cite{Wu:2023}. Our approach expands on this by opening a focus view at each probe location, akin to a 3D portal. We support multiple simultaneous lenses and link them with the main view through continuous cues, thus extending distant object selection into a multi-focus context.

\textbf{Immersive Analytics of Networks}: The application of immersive technology to network analysis has gained traction, demonstrating benefits like improved recall of graph structure and increased engagement \cite{Joos-2025-VNA}. Various application areas ranging from the exploration of biological networks \cite{Pirch:2021} to software engineering have been investigated \cite{Mehra-2019-XRaSE}. However, many existing immersive network tools emphasize exploration and navigation rather than direct editing of the graph \cite{Joos-2025-VNA}, despite being a task in classical graph visualization \cite{Gladisch-2014-SAEG, Eichner-2016-DVEG}. Sorger et al. \cite{Sorger-2019-IAL} used a combination of overview exploration and detail analysis in connection with teleportation. Drogemuller et al.~\cite{Drogemuller-2020-EVR} examined different navigation techniques and found that steering patterns (e.g., one- and two-handed flying) showed benefits for search tasks, while Worlds-in-Miniature approaches better supported overview tasks. Joos et al. \cite{Joos-2024-ENS} compared six distinct node selection techniques, where filtering and volume selections showed advantages for larger graphs. Building on these foundations, our approach introduces volumetric probes that allow users to instantiate multiple localized focus views of the network. These probes serve not only as lenses for inspection but also as portals for interaction, enabling non-local editing operations such as the addition and removal of nodes and links, as well as layout transformation and navigation.

\section{Method}
\label{sec:method}

\begin{figure*}
    \centering
    \begin{subfigure}{0.32\textwidth}
        \begin{tikzpicture}
            \node {\includegraphics[width=\textwidth]{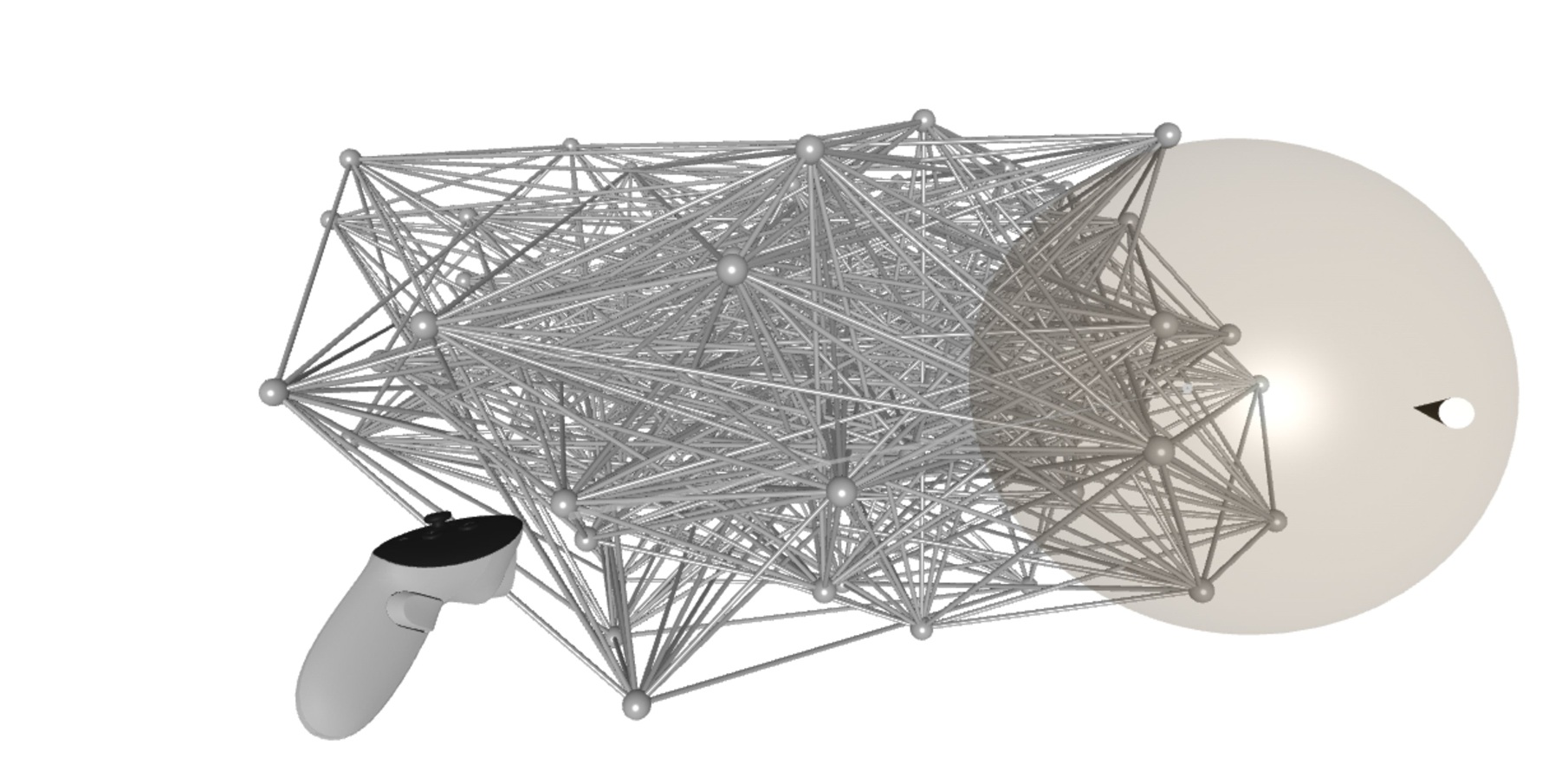}};
            \node at (-1.5,1.2) {\includegraphics[width=0.5\columnwidth]{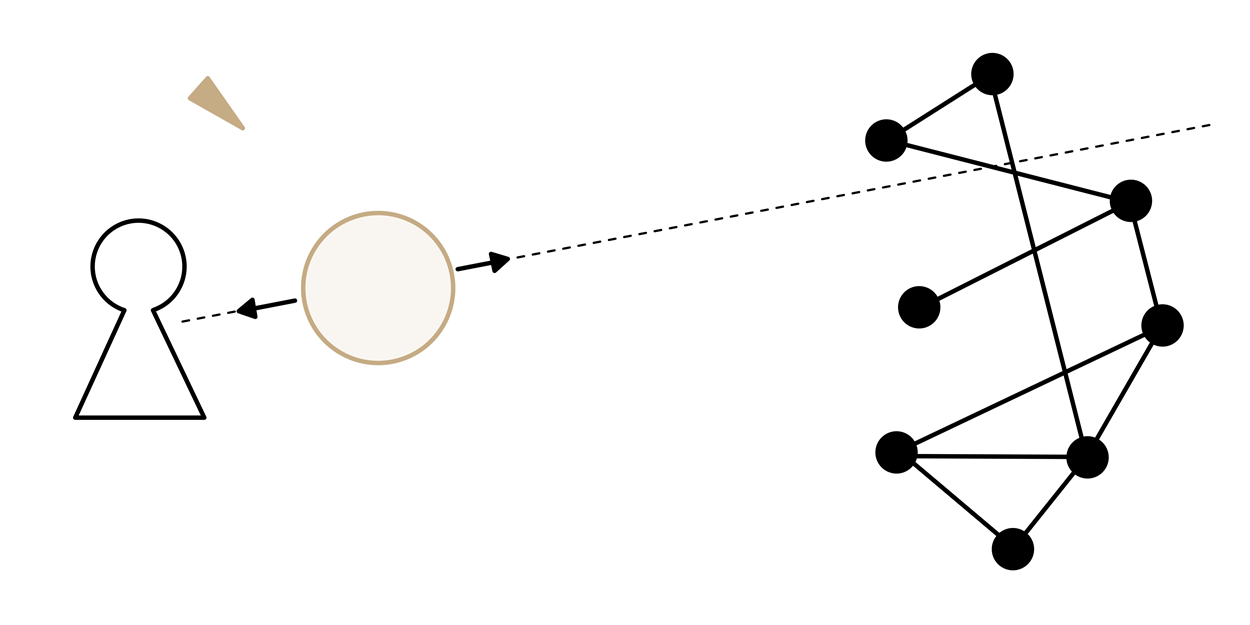}};
        \end{tikzpicture}
        \caption{}
        \label{fig:methodImages_a}
    \end{subfigure}
    ~
    \begin{subfigure}{0.32\textwidth}
    \centering
    \begin{tikzpicture}
        \node {\includegraphics[width=\textwidth]{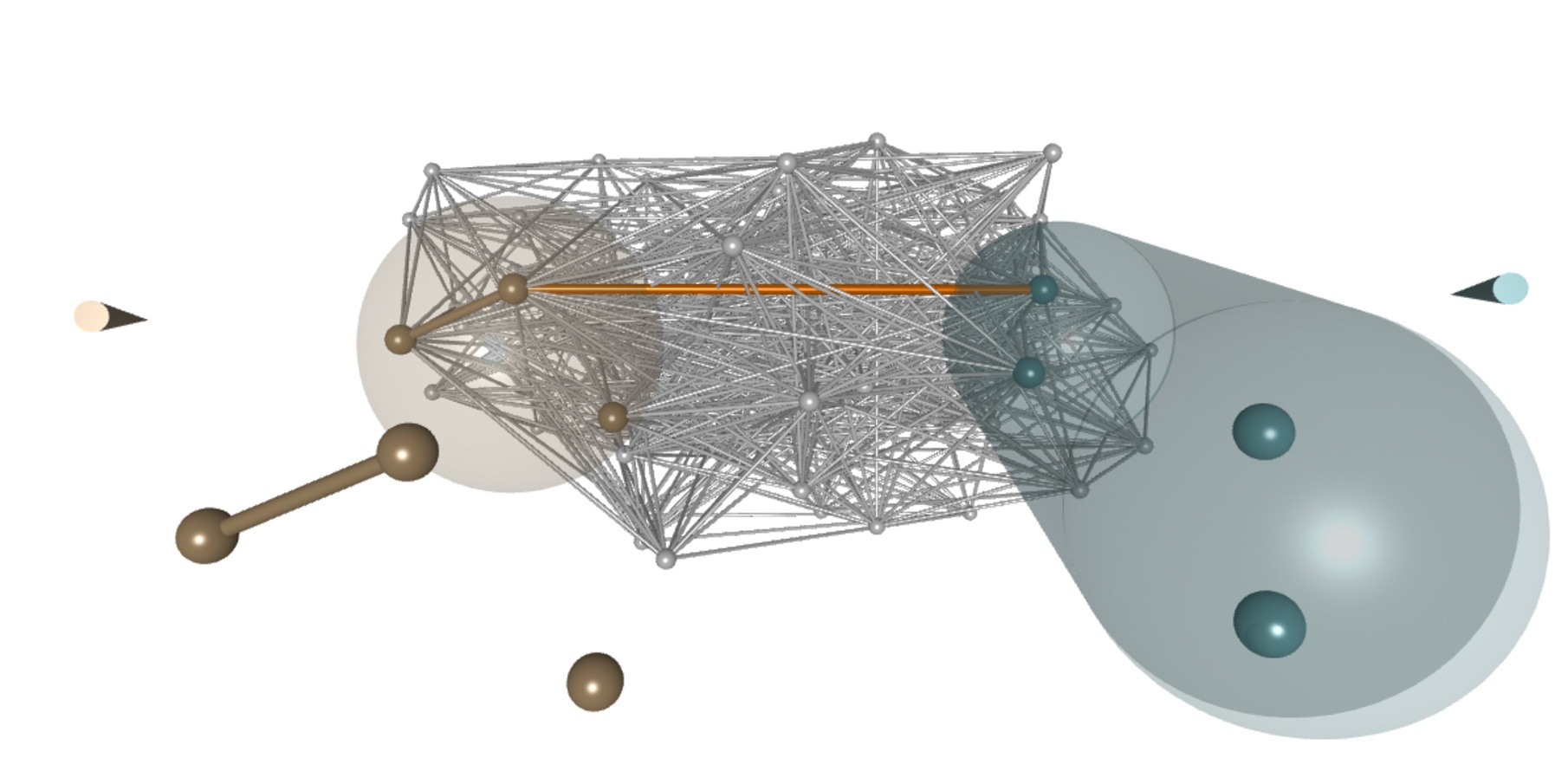}};
        \node at (-1.5,1.2) {\includegraphics[width=0.5\columnwidth]{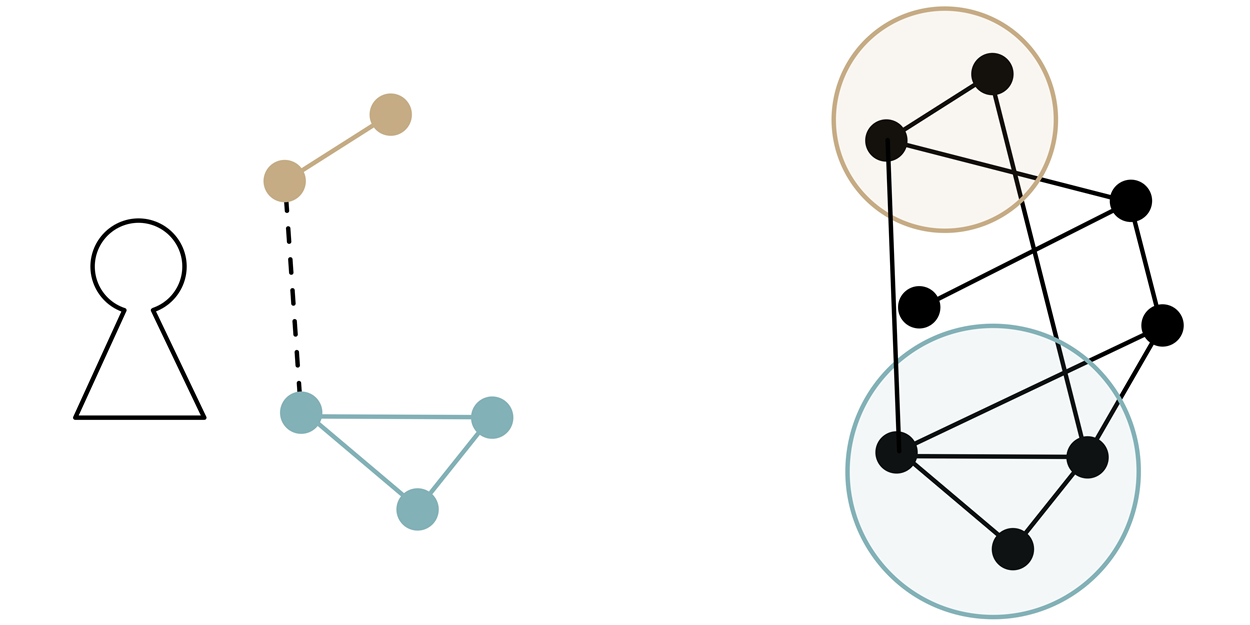}};
    \end{tikzpicture}
    \caption{}
    \label{fig:methodImages_b}
    \end{subfigure}
    ~
    \begin{subfigure}{0.32\textwidth}
    \centering
        \begin{tikzpicture}
            \node {\includegraphics[width=\textwidth]{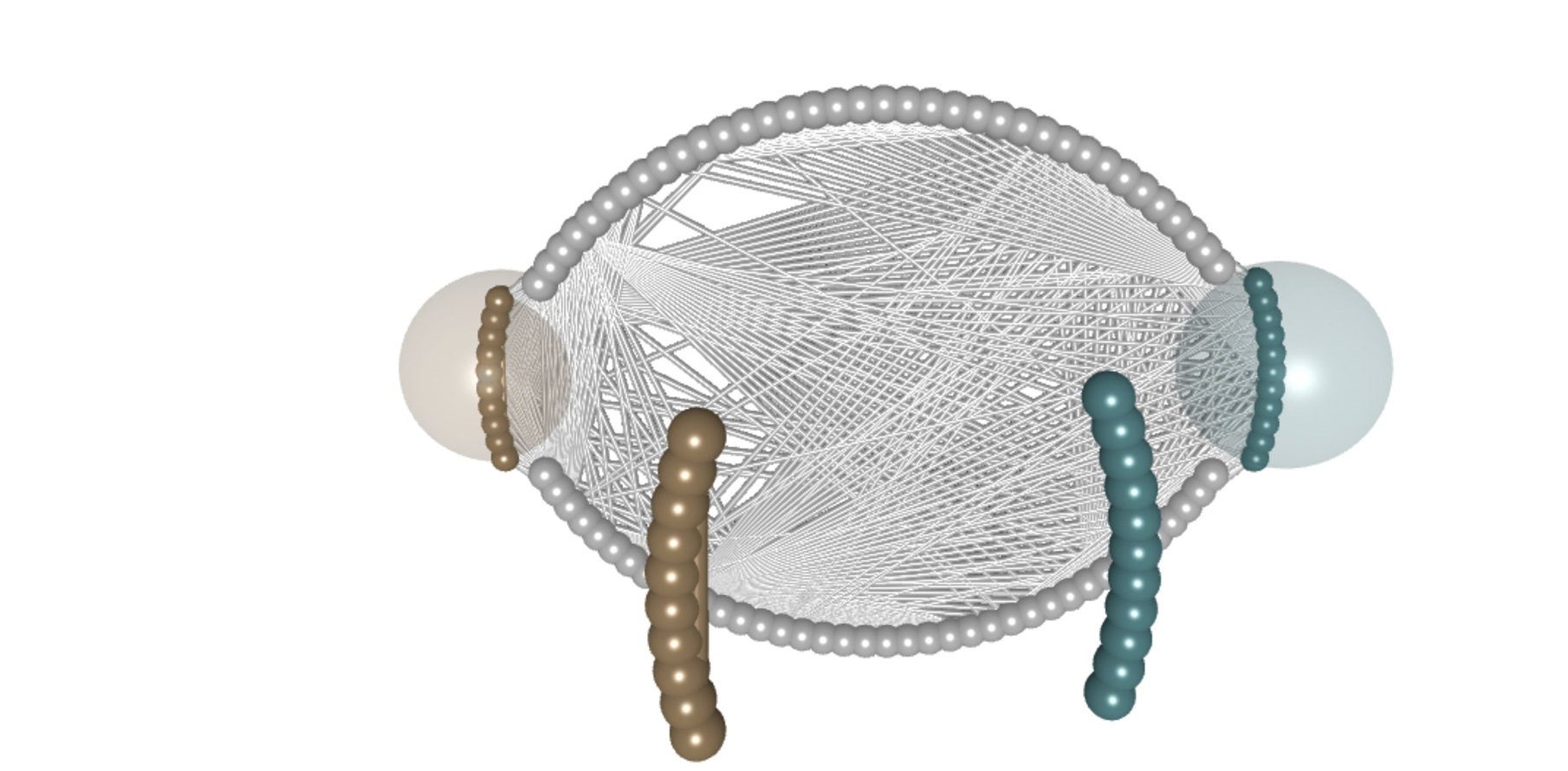}};
            \node at (-1.5,1.2) {\includegraphics[width=0.5\columnwidth]{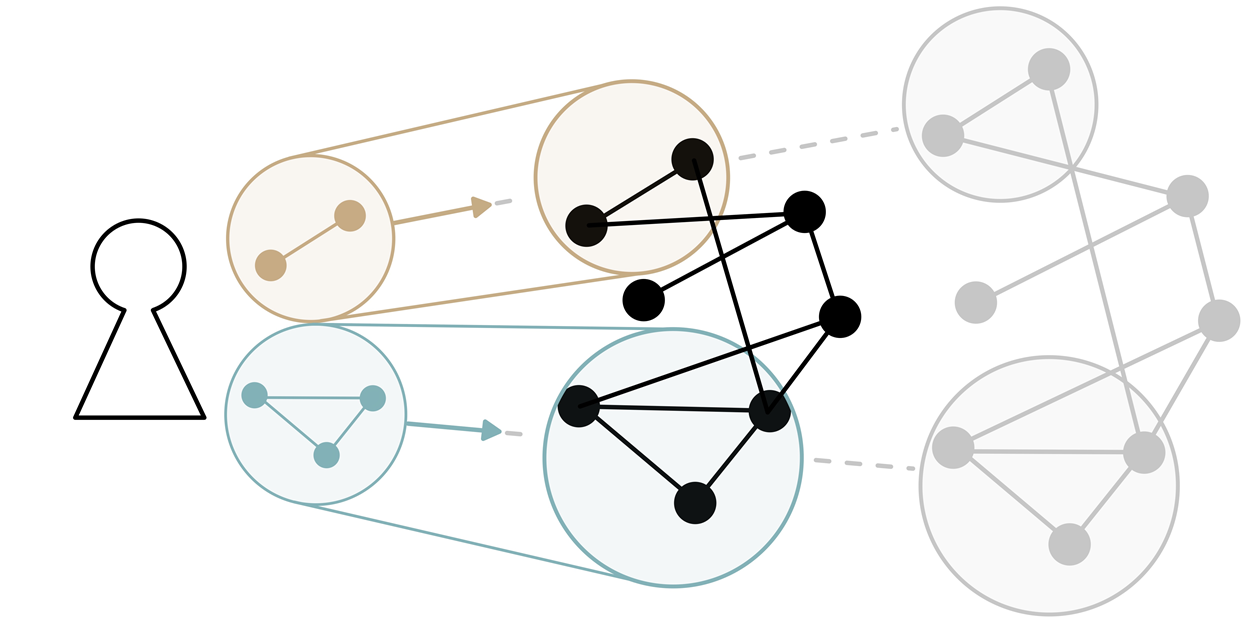}};
        \end{tikzpicture}
        \caption{}
        \label{fig:methodImages_c}
    \end{subfigure}
 \caption{Exemplary renderings and conceptual arts covering the stages explained in \cref{sec:method}. (a) Placement of a multi-focus probe in the graph. (b) Local editing of the probe content while the orange link got created through both subgraphs. Visual cues, i.e., the cones pointing toward the probes and the tunnel (truncated cone) connecting the right probe with its content, support the users' orientation. Additionally, nodes and links in the graph which are surrounded by probes are enlarged and colored with the respective probe color for visibility. (c) Deformation of the graph using two probes. For better clarity, the nodes were initially arranged on a circle and the colored arrows in the concept art indicate the direction vectors caused by the two active probes.}
 \label{fig:methodImages}
\end{figure*}

We integrated Multi-Focus Probes into a VR environment for the exploration and editing of 3D node-link diagrams. Such a diagram, which we also call a \textit{graph}, is a geometric representation of an abstract graph $G = (V, E)$ consisting of a set of vertices $V$ and edges $E$ and it is rendered with spheres and cylindrical tubes representing nodes and links, respectively. Users can navigate the environment using room-scale motion or controller-based input, adopting either an egocentric view (immersed within the graph) or an exocentric view (observing the graph from an external vantage point). In both modes, the goal is to facilitate interaction with arbitrarily positioned subgraphs without requiring extensive physical movement or causing spatial disorientation. Our system supports transitions between egocentric and exocentric perspectives, with the egocentric view shown to improve performance in visual search and navigation tasks while preserving spatial orientation~\cite{Sorger-2021-ENE}. The egocentric perspective is also reflected in the subgraph representations generated by the probes, which act as multiple, independently situated egocentric views. Conceptual illustrations and example renderings of the described components are provided in \cref{fig:methodImages} and in the supplementary video.

\textbf{Placing a Multi-Focus Probe:} To select a region of interest, the user points a VR controller toward the desired location to place a \textit{probe}, cf. \cref{fig:methodImages_a}. We define such a probe as a 3-dimensional closed ball $B(r, b)$ of radius $r \in \mathbb{R}_{>0}$ centered at point $b \in \mathbb{R}^3$ and we represent it by its boundary, i.e., as a semi-transparent, colored 2-sphere. 
Probes are initialized along the controller’s ray, can be interactively scaled and translated along the ray, and positioned freely within the virtual space. Once placed, a probe remains at a fixed location in the global graph. The corresponding focus content, i.e., the subgraph enclosed by the probe, is then automatically extracted and displayed. Existing probes can also be reselected and repositioned or replaced.

\textbf{Manipulating Probe Content:} Once a probe is placed, the system identifies all graph nodes whose centers lie within the probe volume and generates a visualization of the corresponding induced subgraph\footnote{Let $G=(V,E)$ be a graph, $V' \subseteq V$, and $E' \subseteq E$. Then $G' = (V', E') $ is called \textit{subgraph} of $G$. It is called \textit{induced} by $V'$ if every edge in $E'$ has both endpoints in $V'$.} of this intersection in front of the user, colored in the probe's color to indicate its relation to the probe, cf. \cref{fig:methodImages_b}. We call this subgraph the \textit{probe content}. It is visualized as a separate focus view that enables inspection of a localized subset of the overall network. The probe content is positioned relative to the user\textemdash it moves with the user’s viewpoint and can also be freely repositioned. The probe content is fully interactive: users can rotate it, create new links by selecting pairs of nodes, and remove existing nodes or links.  Although each probe operates independently and does not directly interact with other probes, users can manipulate multiple probe contents in parallel. Edits can be applied across different probes or between a probe and the full-scale graph.

\textbf{Global Navigation and Deformation:} 
Our approach supports two global operations that enhance immersive graph exploration: \textit{navigation} toward distant or occluded probes, and \textit{deformation} of the graph layout to reduce visual clutter and disentangle dense structures, cf.~\cref{fig:methodImages_c}. Both are achieved by activating one or more probes, each contributing a directional influence from the probe’s content view (in front of the user) to its origin in the global graph.

Let $\lbrace B_i \rbrace_{i \in I}$ denote the family of active probes for some index set $I$ and $\lbrace v \rbrace_{i\in I}$ the normalized directions from the probes' content center (in front of the user) to its location in the global graph. Then, for a user input received from the controller ranging from -1 to 1, we calculate the scaled directions $\lbrace \tilde{v} \rbrace_{i\in I}$. For each node at position $p$, we determine the set  $J = \lbrace j \in I : p \in B_j \rbrace $ of active probes containing the node. The new position of the node is then computed as:
    \begin{equation}
        p \mapsto p + \frac{1}{\sum_{j\in J} w_j} \sum_{j \in J} w_j \tilde{v}_j
        \label{eq:pointUpdate}
    \end{equation}
using uniform weights $w_j = 1$ if $J \neq \emptyset$. If the node lies outside all active probes (i.e., $J = \emptyset$), we set $J = I$ and use inverse-distance weights $w_j = \left( \norm{p  - b_j} \right)^{-1}$, where $b_j$ is the center of probe $B_j$.

This formulation ensures that nodes enclosed by one or more active probes are displaced according to the averaged directions of those probes, while other nodes are influenced by all probes in proportion to their distance. When only one probe is active, this results in a uniform translation of the entire graph, enabling smooth navigation toward a hidden subgraph (see \cref{fig:connectAndTravel}). When multiple probes are active, their combined influences deform the graph by pulling or pushing different regions in diverging directions (see \cref{fig:deformation}).

In practical terms, probes act like spatial ``handles'' on the graph. Users can pull distant regions toward themselves, stretch apart densely connected areas, or rearrange the structure by coordinating movements across multiple probes. This interaction mechanism provides direct spatial control for resolving occlusion, bringing relevant subgraphs into reach, and improving overall layout clarity in immersive settings.

\textbf{Guidance Cues for Context Preservation:} Introducing multiple focus views and enabling dynamic graph modifications increases the risk of users losing orientation within the data. To mitigate this, we integrate continuous guidance cues (cf. \cref{fig:methodImages_b}). For example, during probe placement, haptic feedback is used: the controller vibrates as long as the probe intersects with at least one node and has not yet been placed, aiding spatial awareness. To support orientation maintenance, we incorporate visual indicators that help users track existing probes, even when they are occluded or positioned at a distance. Each probe is associated with a directional cone~\cite{McCrae:2010}, rendered in the same color as the probe, and pointing toward its location. The cone's transparency encodes the distance to the probe. Let $w$ be the vector from the camera toward the probe and $v$ the current view direction; we compute the angle $\alpha$ between these vectors to determine the cone placement. The cone is positioned by rotating $v$ toward $w$ by a fixed angle and translating it along the resulting vector. To reduce visual clutter, cones are only displayed when $\alpha$ exceeds a defined threshold. A second visual cue, the \textit{tunnel}, is a colored truncated cone connecting each probe to its content. Toggling the tunnel also activates or deactivates the associated probe. Together, these cues support probe localization in dense or spatially distant regions of the graph and provide immediate visual feedback during navigation and deformation.

\section{Results}

\begin{figure*}[tb]
 \centering
 \begin{subfigure}{0.49\textwidth}
 \centering
    \includegraphics[width=0.49\textwidth]{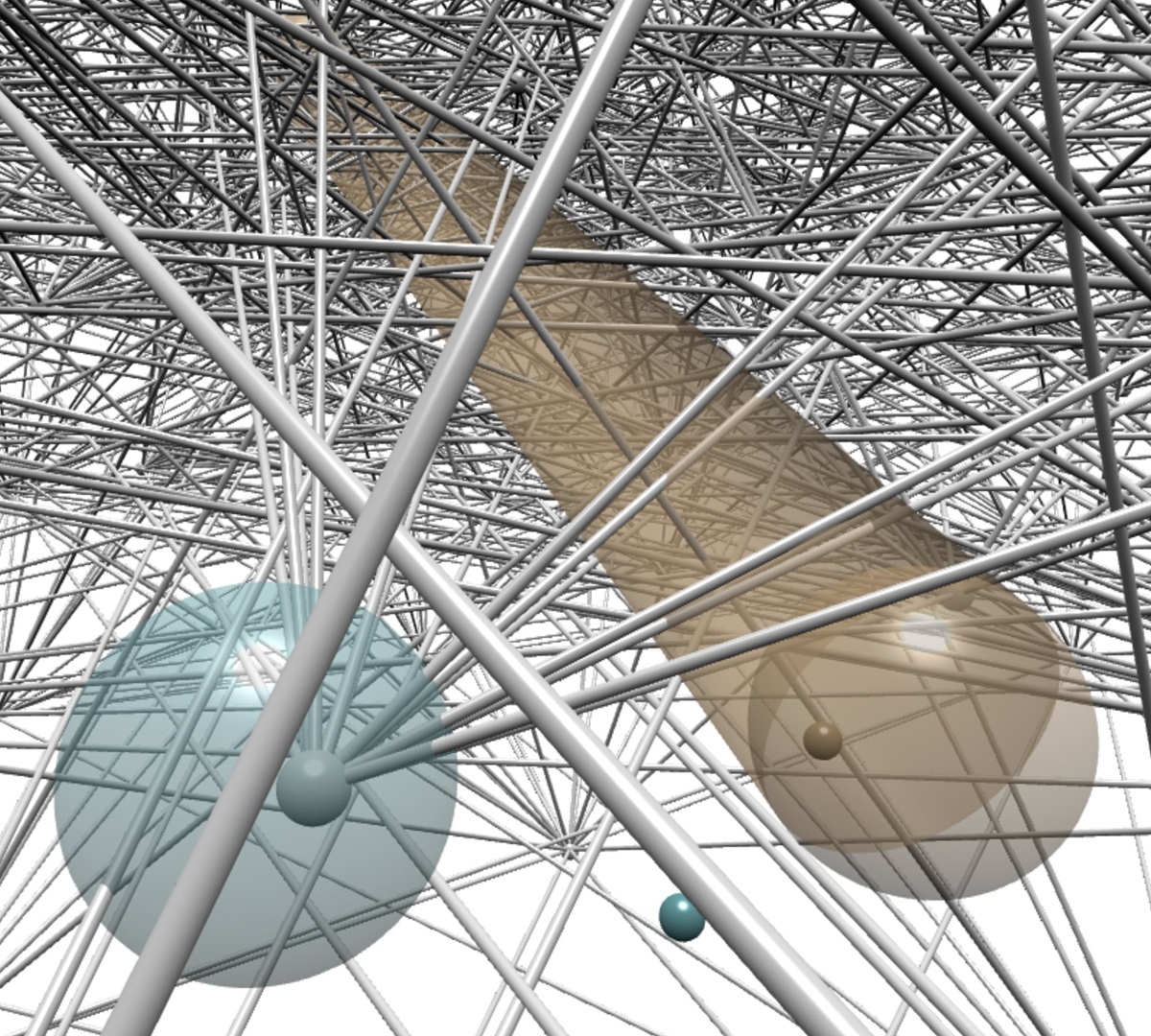}
    \includegraphics[width=0.49\textwidth]{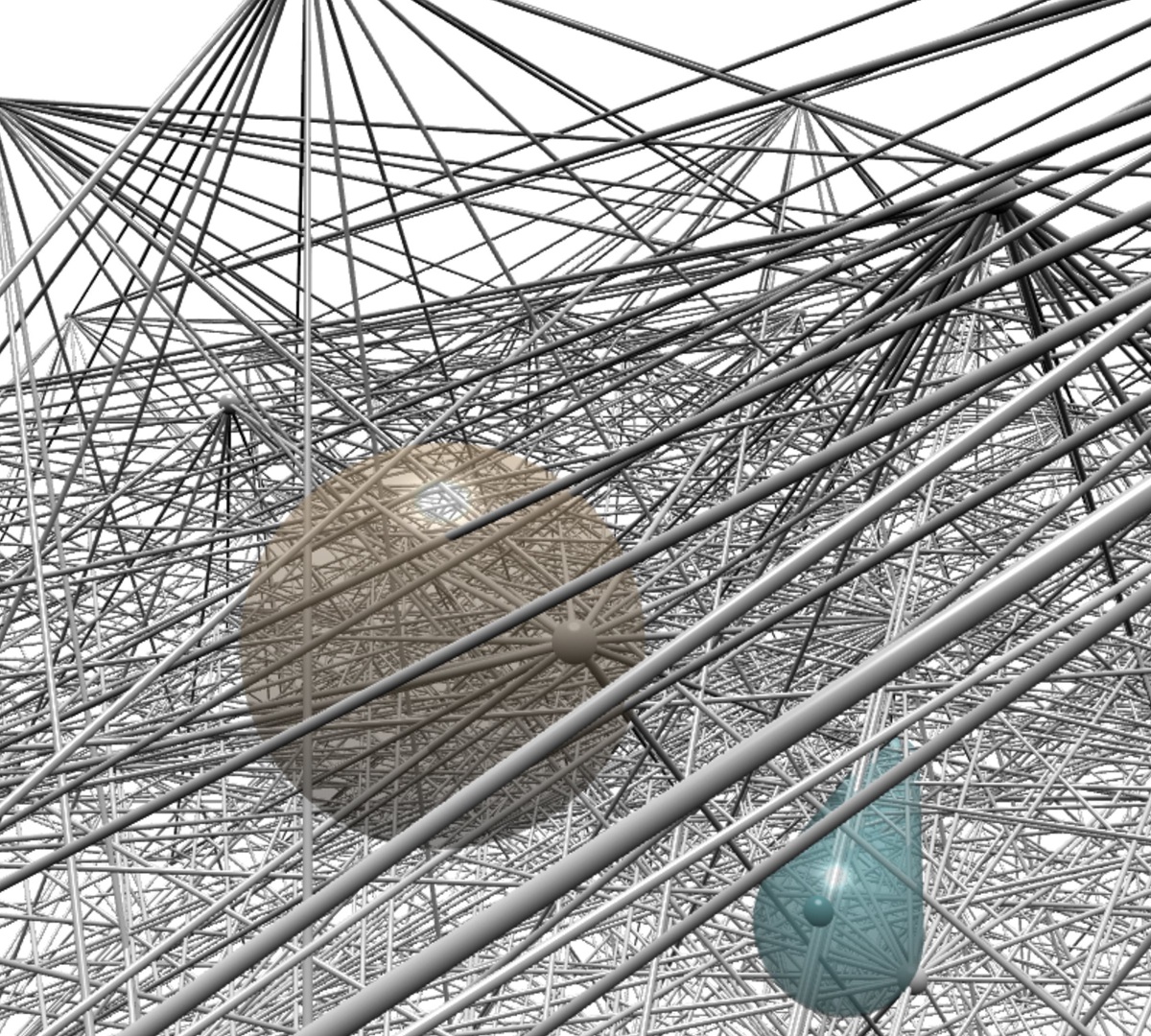}
    \caption{}
    \label{fig:connectAndTravel}
 \end{subfigure}
 ~
\begin{subfigure}{0.49\textwidth}
 \centering
    \includegraphics[width=0.49\textwidth]{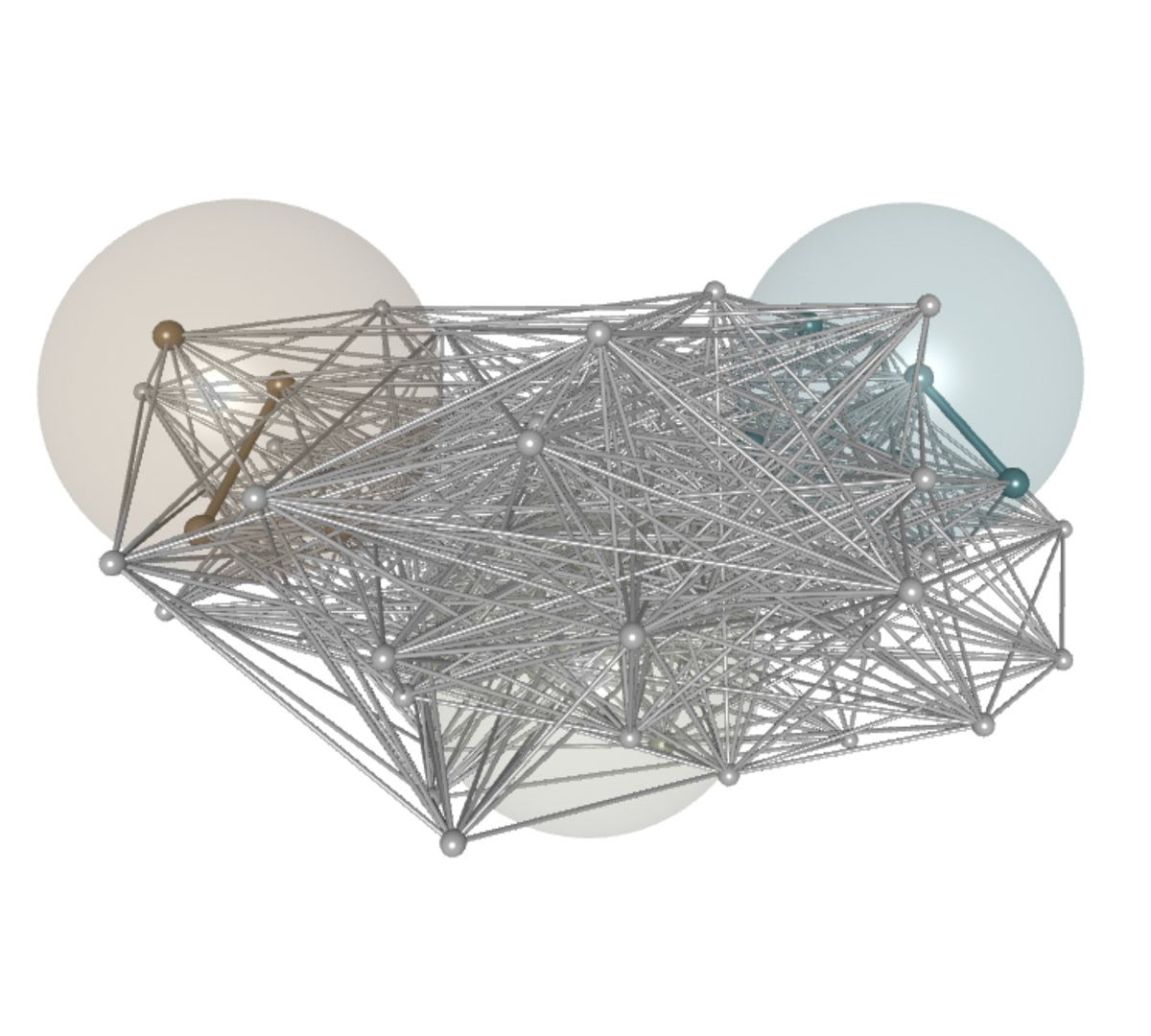}
    \includegraphics[width=0.49\textwidth]{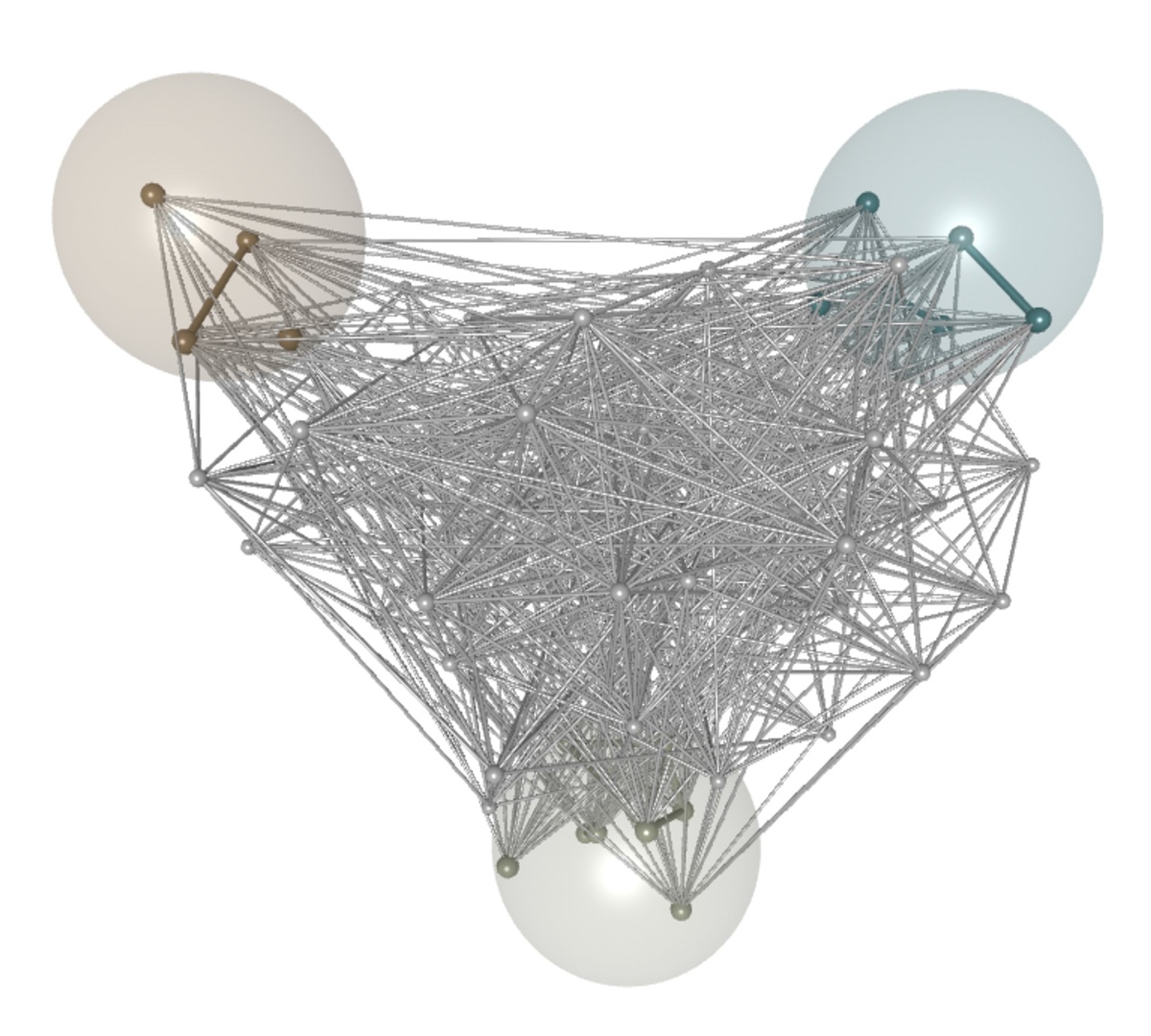}
    \caption{}
    \label{fig:deformation}
 \end{subfigure}
 \caption{(a) Shows viewpoints next to two different probes distant from each other. In each view, the tunnel indicates the path to the respective other probe and whilst only one is active it allows the user to translate to that probe. (b) Depicts the deformation where the left rendering illustrates the initial constellation in which three probes are positioned. On the right the graph is deformed caused by the navigation on one of the probe contents pushing away the probes from each other and stretching the graph in consequence. }
 \label{fig:connectAndTravelAndDeformation}
\end{figure*}

We implemented a prototype for our technique using the web-based 3D engine \textit{babylon.js}\footnote{\url{https://www.babylonjs.com/}}.
In the examples in this paper, we use multivariate graph data representing soccer players from the CL seasons 2017/2018. Each player is represented by a node, with links connecting pairs of players who played for the same club. Additional attributes for each player are, for instance, appearances, minutes played, pass accuracy etc. Overall we have 95 nodes, 1046 links, and 39 node attributes. To obtain an initial embedding of the nodes we used \textit{d3-force-3d}\footnote{\url{https://github.com/vasturiano/d3-force-3d}}, a 3D extension of the physical force simulation on particles included in D3\footnote{\url{https://d3js.org/}}. We initialize the scene by placing the rescaled graph in front of but distant from the user.


In \cref{fig:teaser}, the graph is shown rendered in front of the user, with two probes placed and activated. This setup presents two primary challenges. The first is the distance between the user and the graph, which makes precise manual probe placement difficult. To address this, haptic feedback supports depth perception during interaction. The second challenge lies in the dense entanglement of graph elements. Here, users can progressively insert a probe into the graph to explore occluded regions. Alternatively, probes can be placed automatically based on node attributes or structural properties of the graph. This extends the interaction capabilities by enabling selection of objects through high-level queries—for instance, filtering for a node representing the player with the highest number of minutes played. Since automatically placed probes might be outside the user's field of view or hidden within the graph, we employ visual cues (such as cones and tunnels) to guide the user’s attention and provide spatial orientation.

The content of a probe, positioned relative to the user, serves as an interaction proxy that bridges the spatial gap between the user and specific subsets of the graph. This setup enables users to explore content and perform local edits\textemdash either within a single subgraph, across multiple subgraphs, or in combination between subgraphs and the main graph. Current editing capabilities include creating nodes and removing nodes or links. In \cref{fig:teaser}, for instance, two distant nodes are connected via a newly created link, resulting from the selection of these nodes in their respective subgraphs.

Since the probe content serves as a connection to the graph, it also carries system control functionality. This includes toggling the probe’s activation state, which in turn controls the visibility of the tunnel used as a visual guidance cue. Further, navigation and deformation can be initiated and controlled via the probe content.

As recalled from \cref{eq:pointUpdate}, a single active probe enables scene navigation relative to the probe, drawing the user toward or pushing them away from its position. \cref{fig:connectAndTravel} shows two egocentric views taken near each of two probes. When multiple probes are active, they can be used for deformation. For example, in \cref{fig:deformation}, three probes were employed to stretch the graph. Here, replacing the probe content in front of the user allows them to steer the deformation, as the corresponding translation vectors are computed from the probe content’s position to the respective probe locations.

This type of deformation can be used to bring nodes of interest, especially those belonging to different subgraphs, closer to the user and to each other, while maintaining reasonable behavior for the remaining nodes. In other scenarios, dense entanglements may hinder the exploration of interrelationships. In such cases, moving probe content and thereby stretching the graph helps to unravel tightly clustered nodes and links (cf. \cref{fig:deformation}).

\section{Discussion}

Our current implementation serves as a proof of concept for the multi-focus probe technique and includes basic interaction and editing functionality. A key direction for future work is a thorough empirical evaluation to validate usability and effectiveness. This begins with the challenge of probe placement in dense or distant regions of the graph, an inherently difficult task, even with haptic feedback. To improve this interaction, we plan to evaluate alternative selection strategies, including multimodal feedback, automated placement, and probe-linked cameras. For instance, coupling probe-based navigation with automatic probe initialization may benefit exploration and search tasks. We also hypothesize that direct, embodied interaction with subgraph representatives\textemdash especially when distributed across multiple locations in the graph\textemdash offers advantages such as reduced spatial disorientation, more efficient task-switching between regions of interest, and the ability to perform comparative or coordinated edits without the need for continuous navigation or viewpoint changes. Furthermore, the local (manual) arrangement, i.e., positioning, of probe contents, relating to multi-view layouts \cite{Satriadi-2020-MAM}, can be extended to steer the deformation process even further. Currently, the direction vectors (cf. deformation in \cref{sec:method}) are employed and the positioning of contents in front of the user influence the deformation. Thus, including additional entities like direction vectors obtained from inter-relations of probe contents could provide a mechanism for which the user gains more fine-grained control of the whole deformation mainly based on how they locally arrange the contents w.r.t. each other.

Another important direction for future work lies in expanding the interaction vocabulary, also connected to relevant tasks \cite{Gladisch-2015-TIGEE}. While the current implementation focuses on basic editing and deformation, additional probe-based operations such as attribute-based filtering, multi-probe coordination, and probe history tracking could further increase analytical expressiveness. Multi-focus probes may also serve as an mechanism for collaborative scenarios. In a shared immersive environment, probes could be used not only for personal exploration but also as a means of communicating and sharing focus views with others. For instance, users could expose their probe contents to collaborators to jointly inspect, annotate, or manipulate specific subgraphs. 

From a technical perspective, egocentric views \cref{fig:connectAndTravel} pose challenges in dense graphs, particularly for the visual organization of probe contents, guidance cues, and the graph structure itself. Interactive elements may become occluded by nodes or links, limiting usability. In this context, distortion-based viewing techniques in 3D space~\cite{Carpendale:1996} could be explored to enhance spatial awareness and operability in egocentric perspectives. Additionally, the rendering of links, especially w.r.t. applied graph deformations, complicates or even hinders the topological perception of the graph. To address this, alternative rendering techniques like curved links \cite{Yang-2019-FMIE}, edge bundling \cite{Kwon-2016-LRIIGV}, or the representation of links by curved tubes with varying radii could be incorporated to support structural perception and the overall understandability. Moreover, our current work targets graphs of moderate sizes. Scaling the technique to handle, e.g., genome-scale networks with tens of thousands of nodes and hundreds of thousands of links [19], presents both a technical challenge and an opportunity to broaden the range of application scenarios. Beyond graph data, our method could also be extended to other dense structures such as molecular or volumetric data, where the concept naturally aligns with endoscopic-style exploration.

\section{Conclusion}
In this work, we introduced the multi-focus probes enabling interaction with visualizations of complex and/or dense graphs in virtual environments. We demonstrated how these probes support local editing, focus-based navigation, and graph deformation. To enhance usability, we incorporated visual cues aligned with the focus+context paradigm and enabled seamless switching between exocentric and egocentric perspectives. While our focus on editing marks an important first step, further research is needed to bring immersive environments closer to the level of power and flexibility for network manipulation that is currently available in advanced 2D desktop applications.

\newpage

\bibliographystyle{abbrv-doi}

\bibliography{article}
\end{document}